\documentclass{PoS}

\title{Transient gamma-ray emission from Cygnus X-3}

\ShortTitle{Transient gamma-ray emission from Cygnus X-3}

\author{Anabella T. Araudo %\thanks{}
\\
Instituto Argentino de
Radioastronom\'{\i}a, C.C.5, (1894) Villa Elisa, Buenos Aires,
Argentina. \\
Facultad de Ciencias Astron\'omicas y Geof\'{\i}sicas,
Universidad Nacional de La Plata, Paseo del Bosque, 1900 La Plata,
Argentina. \\
        E-mail: \email{aaraudo@fcaglp.unlp.edu.ar}}

\author{Valent\'{\i} Bosch-Ramon\\
Dublin Institute for Advanced Studies, 31 Fitzwilliam Place, Dublin 2.\\
        E-mail: \email{valenti@cp.dias.ie}}

\author{Gustavo E. Romero\\
Instituto Argentino de
Radioastronom\'{\i}a, C.C.5, (1894) Villa Elisa, Buenos Aires,
Argentina. \\
Facultad de Ciencias Astron\'omicas y Geof\'{\i}sicas,
Universidad Nacional de La Plata, Paseo del Bosque, 1900 La Plata,
Argentina. \\
        E-mail: \email{romero@fcaglp.unlp.edu.ar}}

\abstract{The high-mass microquasar Cygnus X-3 has
been recently detected in a flaring state by the gamma-ray satellites
Fermi and Agile. In the present  contribution, we study the high-energy
emission from Cygnus X-3 through a model based on the interaction of
clumps from the Wolf-Rayet wind with the jet. The clumps inside the jet act as
obstacles in which shocks are formed leading to particle acceleration
and non-thermal emission. We model the high energy emission  produced
by the interaction of one clump with the jet and briefly discus the possibility
of many clumps interacting with the jet. From the characteristics of
the considered scenario, the produced emission could be flare-like due
to discontinuous clump penetration, with the GeV long-term activity
explained by changes in the wind properties.}

\FullConference{25th Texas Symposium on Relativistic Astrophysics - TEXAS 2010\\
		December 06-10, 2010\\
		Heidelberg, Germany}

\begin{document}

\section{Introduction} 

\begin{figure}
\includegraphics[angle=0, width=0.5\textwidth]{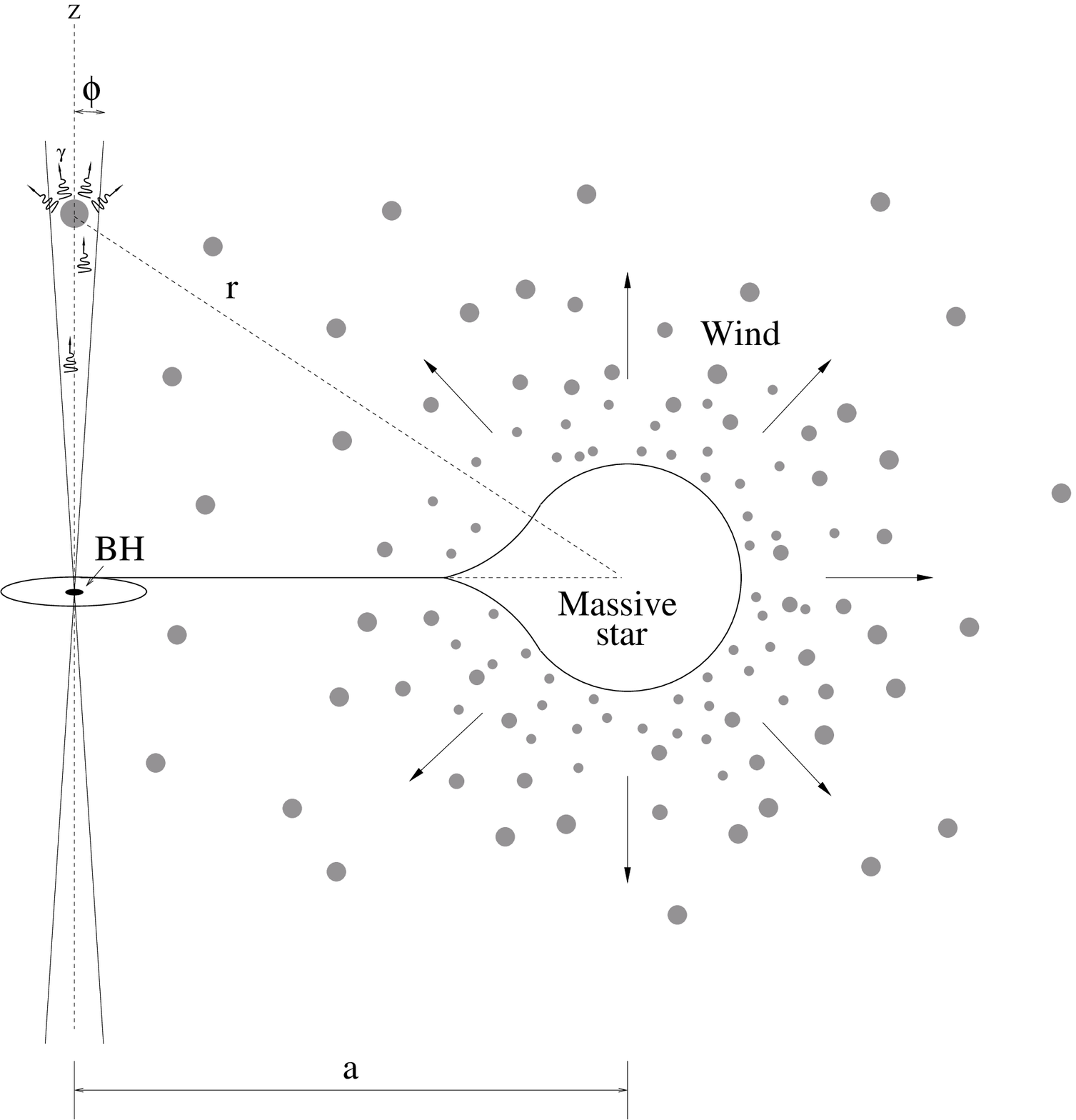}
\includegraphics[angle=0, width=0.4\textwidth]{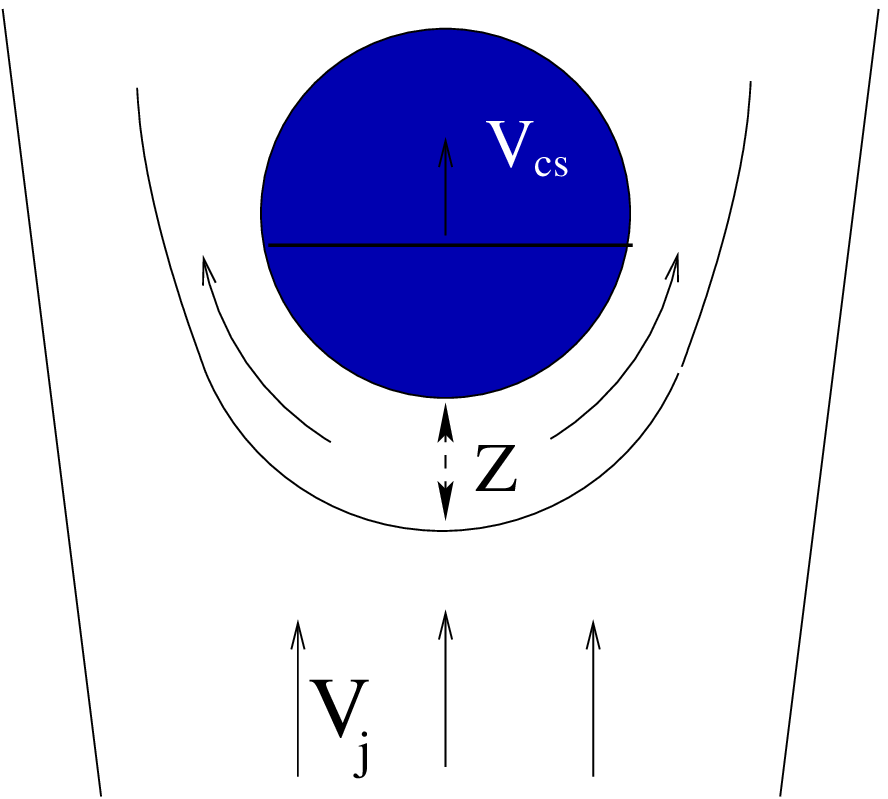}
\caption{Left: Sketch of a HMMQ. Right: Sketch of the
interaction of one clump with the jet.}
\label{scenario}
\end{figure}

The mass loss in massive stars is
thought to take place via supersonic inhomogeneous winds. 
Considerable observational evidence supports the idea that the wind
structure is clumpy~\cite{owocki_cohen}, 
although the properties of clumps are not 
well-known as a consequence of the very high spatial resolution necessary for 
clump detection. 
Some massive stars are accompanied by a compact object
and present transfer of matter to the latter, forming
an accretion disc. When bipolar relativistic outflows are formed, these sources
are called high-mass microquasars (HMMQs). In Figure~\ref{scenario} (left),  
a sketch of a HMMQ is shown. 

Cygnus~X-3 is a HMMQ composed by a Wolf-Rayet (WR) star
and a compact object, both being separated a distance
$a \sim 3\times10^{11}$~cm. The nature of the compact object has not been
determined yet, but recent studies suggest that it may be a black 
hole (e.g. \cite{shrader}).  
Located at a distance $\sim 7-9$~kpc, the system
has been recently detected in a recurrent flaring state
at high-energy (HE) gamma rays~\cite{fermi,agile}. The observed emission
spectrum is a power-law with index $\sim 1.7$ and 
luminosity $\gtrsim 10^{36}$~erg~s$^{-1}$. 
The typical duration of the flares was $\sim 2$~days~\cite{agile}.

In this contribution, we present a model to explain the %production of
HE emission detected in Cygnus~X-3  based on the interaction between 
its jets and wind inhomogeneities~\cite{jet-clump}. 
Some clumps can eventually penetrate in the jet, 
leading to transient non-thermal activity that may
release a significant fraction of the jet kinetic luminosity, $L_{\rm j}$, 
in the form of synchrotron, inverse Compton (IC), and proton-proton ($pp$) 
emission.

\section{Jet-clump interaction in Cygnus X-3}

In order to study the physical processes of the jet-clump interaction, 
we consider the collision of  a single clump with the jet, as is shown
in Figure~\ref{scenario} (right).
A huge density contrast $\chi \equiv \rho_{\rm c}/\rho_{\rm j}$ 
(where $\rho_{\rm c}$ and
$\rho_{\rm j}$ are the clump and jet densities, respectively) 
 allows the clump to fully cross the
boundary of the jet and penetrate into it. 
In the context of this work, thermal conduction, magnetic
fields and gravitational forces are not dynamically relevant for the
jet-clump interaction and have been neglected. 

We consider that a clump with size 
$R_{\rm c}^0 = 10^{10}$~cm ($\lesssim 0.1 R_{\star}$)
reaches one of the HMMQ jets moving at the terminal wind velocity
$v_{\infty} = 2500$~km~s$^{-1}$. At a distance $r = \sqrt{z^2 + a^2}$ from
the star, the clump density can be estimated  
through $\rho_{\rm c} = \rho_{\rm w}/f$, where $f < 1$ is the
filling factor of the clumpy wind, and 
$\rho_{\rm w} \sim \dot M/(4\pi r^2 v_{\infty})$ is the wind density,
where $\dot M \sim 10^{-5}\,{\rm M_{\odot}\,yr^{-1}}$ is the typical 
mass loss rate of WR stars.
We assume that the jet has a  velocity $v_{\rm j} \sim c/3$
(i.e. Lorentz factor $\Gamma \sim 1.06$) 
and a kinetic luminosity $L_{\rm j} = 10^{38}$~erg~s$^{-1}$. We fix
the relation between the height (z) and the width ($R_{\rm j}$) of the jet as 
$R_{\rm j} = 0.1z$.    

\begin{figure}
\includegraphics[angle=270, width=0.49\textwidth]{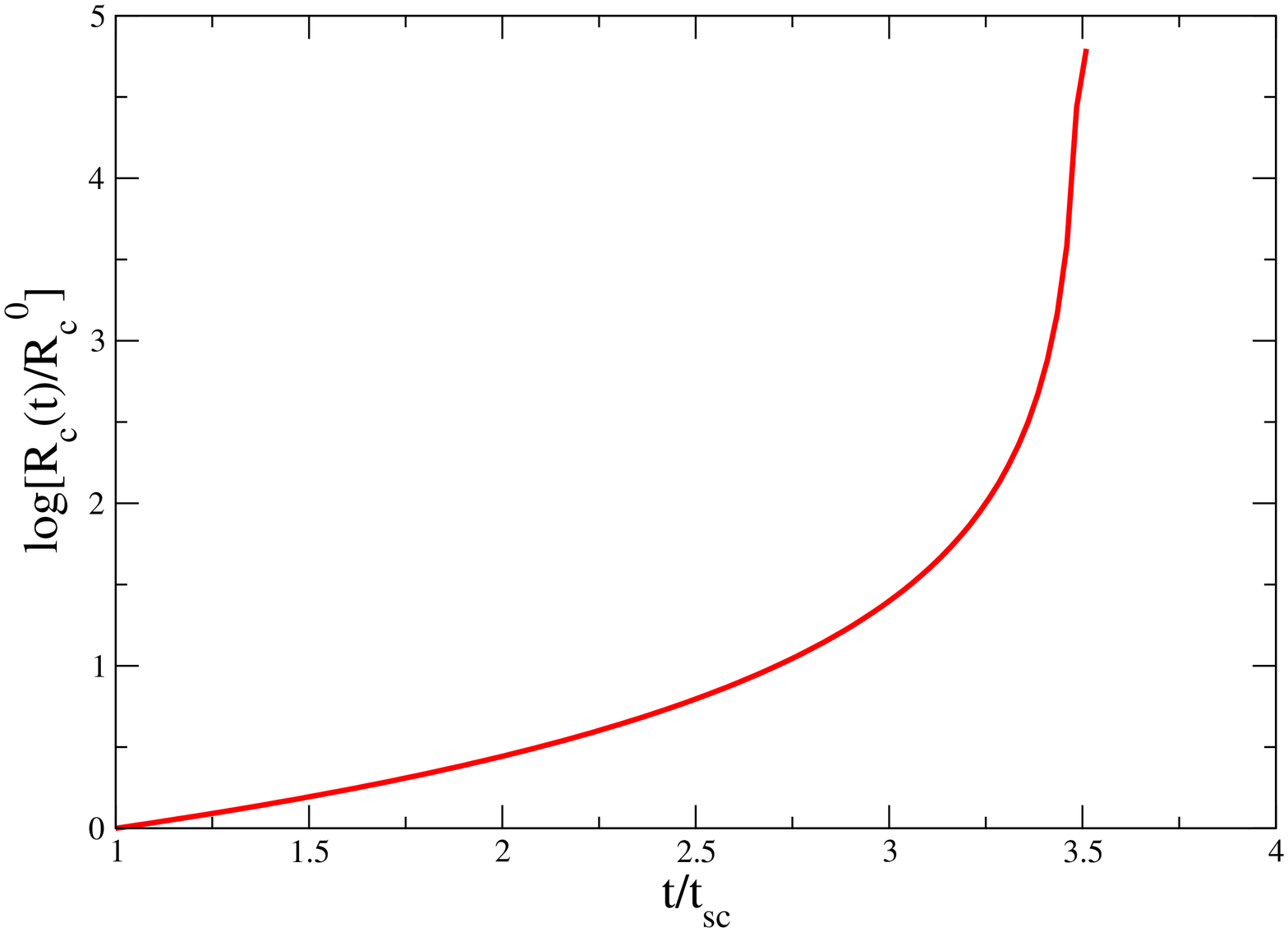}
\includegraphics[angle=270, width=0.49\textwidth]{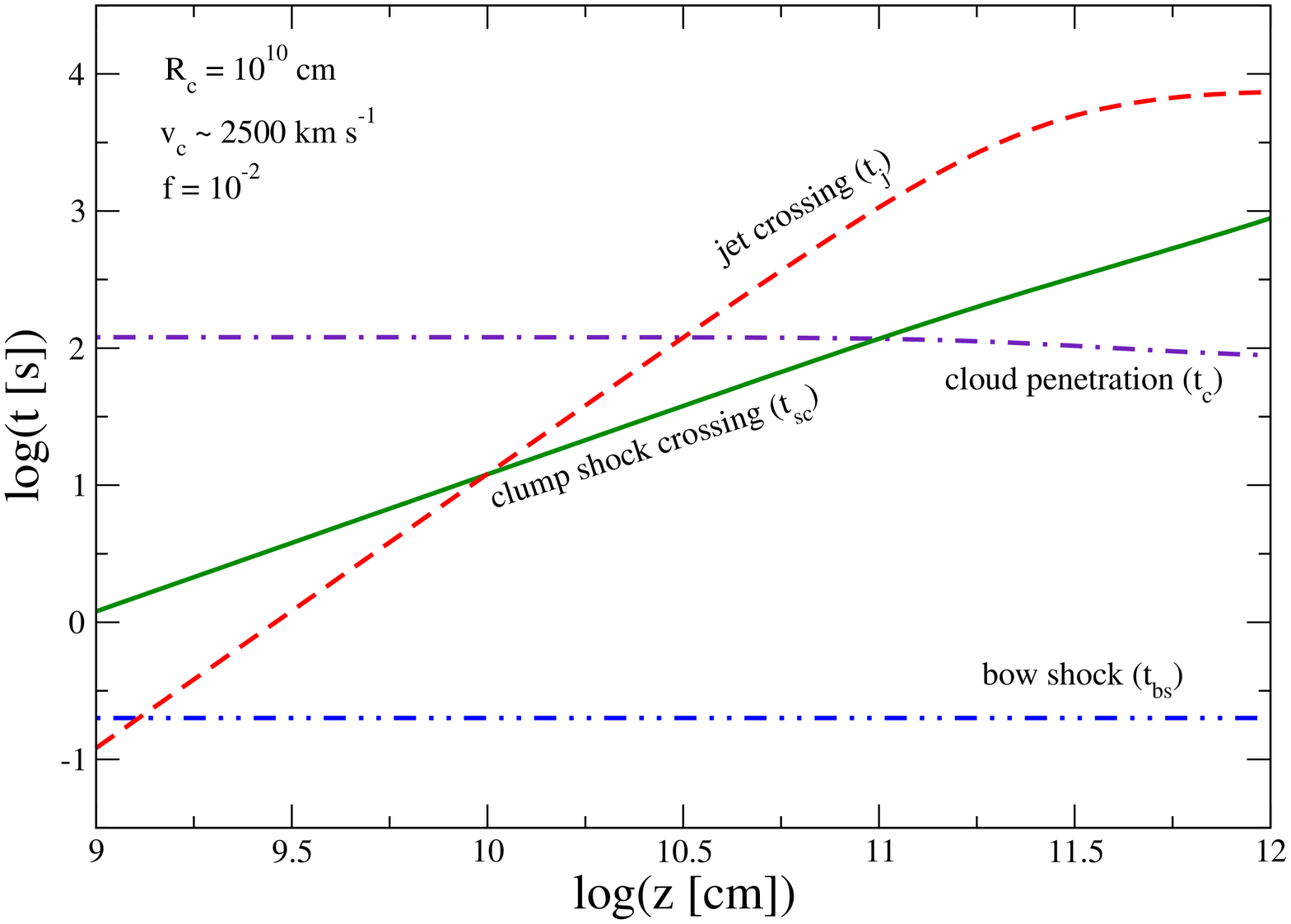}
\caption{Left: Clump spherical expansion. 
Right: Dynamical timescales corresponding to the interaction of one clump 
of $R_{\rm c}^0 = 10^{10}$~cm with the jet.}
\label{exp}
\end{figure}

The penetration time of the clump into the jet is determined by 
$t_{\rm c} \sim 2 R_{\rm c}^0/v_{\rm c} \sim 100$~s, 
where the clump velocity is $v_{\rm c}= v_{\infty}$.  
As a consequence of the interaction of the jet material with the clump, 
two shocks form.
One of these shocks propagates back in the jet 
with a velocity $v_{\rm bs} \sim v_{\rm j}$, forming a bow shock 
(see Figure~\ref{scenario}, right).
This bow shock reaches the steady state configuration  in a time 
$t_{\rm bs} \sim Z_{\rm bs}/v_{\rm bs}$, with the stagnation point 
located at a distance $Z_{\rm bs} \sim R_{\rm c}^0/5$ from the clump.  
On the other hand, a shock propagates in the clump at a velocity
$v_{\rm sc} \sim v_{\rm j}/\sqrt{\chi}$
and, in a time $t_{\rm sc} \sim 2R_{\rm c}/v_{\rm sc}$, the whole 
clump is shocked. 
During the shock passage, the clump is compressed in the direction of
the shock velocity, $z$, and when the whole clump is 
shocked, its size in the $z$-direction is $\sim R_{\rm c}^0/4$. 
However, in $t >  t_{\rm sc}$ the clump starts to expand
at the sound velocity of the shocked material. At the same time, the 
acceleration exerted by the jet to the clump increases with $R_{\rm c}$.   
Assuming that the expansion is spherical, and neglecting the
acceleration exerted by the jet on the clump,
we obtain the following expression for the
clump size: $R_{\rm c}(t) \sim R_{\rm c}^0/(1 - 0.4t/t_{\rm sc})^2$~\cite{maxim}, 
which gives a characteristic clump expansion time 
$t_{\rm exp} \sim 3.5t_{\rm sc}$. 
In Figure~\ref{exp}~(left), the evolution of $R_{\rm c}$ with time is shown.
 
As a consequence of the acceleration exerted by the jet on the clump, 
Rayleigh-Taylor (RT) instabilities can develop in the contact discontinuity. 
Kelvin-Helmholtz (KH) instability can also grow as a 
result of the high relative velocity between the jet
shocked material and the clump.
Numerical simulations show that RT and KH instabilities
grow sufficiently to destroy the clump (i.e.  up to a wavelength 
$\sim R_{\rm c}$) in a timescale 
$t_{\rm RT} \sim t_{\rm KH} \sim 5 t_{\rm sc}$~\cite{klein}. 
If instabilities do not grow fast enough, and clump expansion and acceleration
are too slow,
the permanence of the clump into the jet will be determined by the passage 
time of the clump through the jet: $t_{\rm j} \sim 2R_{\rm j}/v_{\rm c}$.
In Figure~\ref{exp}~(right), the main dynamical timescales are plotted.

\subsection{Interaction height}

Considering the previous analysis on the dynamics of the
jet-clump interaction, one can determine the minimum interaction height, 
$z_{\rm int}^{\rm min}$, at which
one clump can completely penetrate into the jet. Being the lifetime of 
clumps into the jet $t_{\rm life} \gtrsim t_{\rm sc}$, to determine
the interaction height $z_{\rm int}$ we impose that $t_{\rm sc} > t_{\rm c}$. 
This constraint is fulfilled if  $z > A \sqrt{z^2 + a^2}$, where 
\begin{equation}
\label{int}
A \sim 0.5 \left[\left(\frac{f}{0.01}\right)
\left(\frac{v_{\rm j}}{c/3}\right)
\left(\frac{L_{\rm j}}{10^{38}~{\rm erg/s}}\right)\right]^{1/2}
\left[\left(\frac{\Gamma_{\rm j}-1}{0.06}\right)
\left(\frac{\dot M}{10^{-5}~M_{\odot}/{\rm yr}}\right)
\left(\frac{v_{\infty}}{2500~{\rm km/s}}\right)\right]^{-1/2}.
\end{equation}
If, for a  certain election of clump and jet parameters, $A$ results larger 
than 1, the constrain  $t_{\rm sc} > t_{\rm c}$ is no satisfied for any value of
$z$ and clumps will
be  destroyed by the jet before fully penetrating  into it.   
However, for the parameters chosen in this work, $A \sim 0.5$ and in 
such a case, 
the minimum interaction height results 
$z_{\rm int}^{\rm min} \sim A\,a/\sqrt{1 - A^2} \sim a/2$.

\section{Non-thermal processes}

To estimate the non-thermal emission produced by the 
interaction of one clump with the jet, we fix the interaction height 
at $z_{\rm int} = a \sim 2\,z_{\rm int}^{\rm min}$. At $z_{\rm int}$, the density
of the clump is $\rho_{\rm c} \sim 10^{-12}$~gr~cm$^{-3}$ (i.e. the number
 density is $n_{\rm c} = \rho_{\rm c}/(4 m_p) \sim 2\times10^{11}$~cm$^{-3}$,
where $m_p$ is the proton mass).

\subsection{Particle acceleration}

In the bow shock, we assume that particles are accelerated up to 
relativistic energies
being injected in the downstream region following a distribution
$Q_{e,p} \propto E_{e,p}^{-2}$ ($e$ and $p$ for electrons and protons, 
respectively) and with a luminosity 
$L_{\rm nt} = \eta_{\rm nt}(R_{\rm c}/R_{\rm j})^2L_{\rm j}$, where 
we fix $\eta_{\rm nt} = 0.1$. For the parameters assumed in this work,
\begin{equation}
L_{\rm nt} \sim 10^{36}\left(\frac{\eta_{\rm nt}}{0.1}\right)
\left(\frac{R_{\rm c}^0}{10^{10}\,{\rm cm}}\right)^2
\left(\frac{z_{\rm int}}{a = 3\times10^{11}\,{\rm cm}}\right)^{-2}
\left(\frac{L_{\rm j}}{10^{38}\,{\rm erg\,s^{-1}}}\right)
\qquad{\rm erg~s^{-1}}.
\end{equation}
In addition, 
considering that the magnetic energy density in the bow-shock region, $U_{B}$,
is a fraction $\eta_{\rm B}$ of the non-thermal one, $U_{\rm nt}$, we obtain 
a magnetic field $B \sim \sqrt{\eta_{\rm B}}\,2\times10^3$~G. (Both
$\eta_{\rm nt}$ and  $\eta_{\rm B}$ are free parameters in our work.)

\begin{figure}
\includegraphics[angle=270, width=0.49\textwidth]{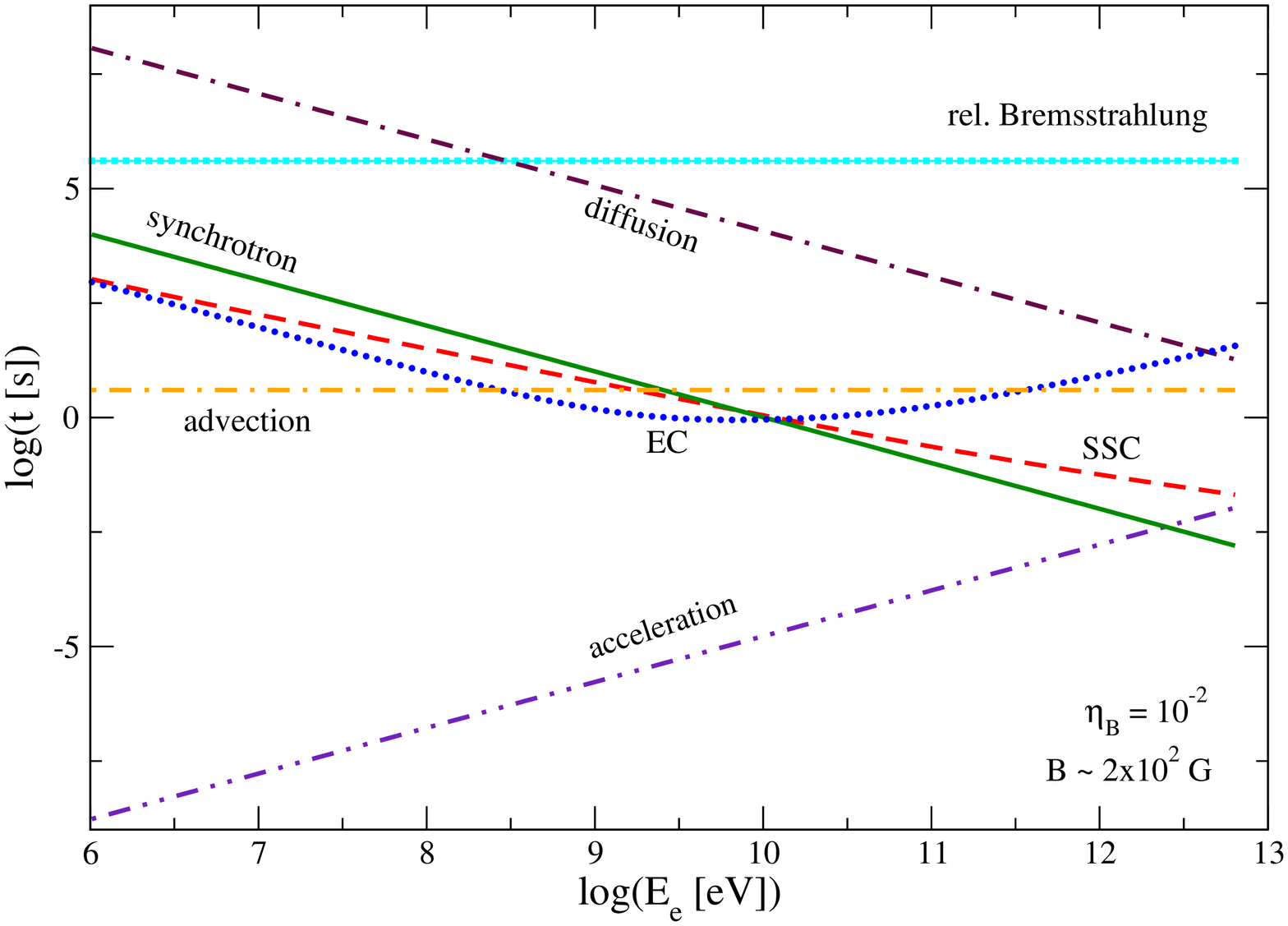}
\includegraphics[angle=270, width=0.49\textwidth]{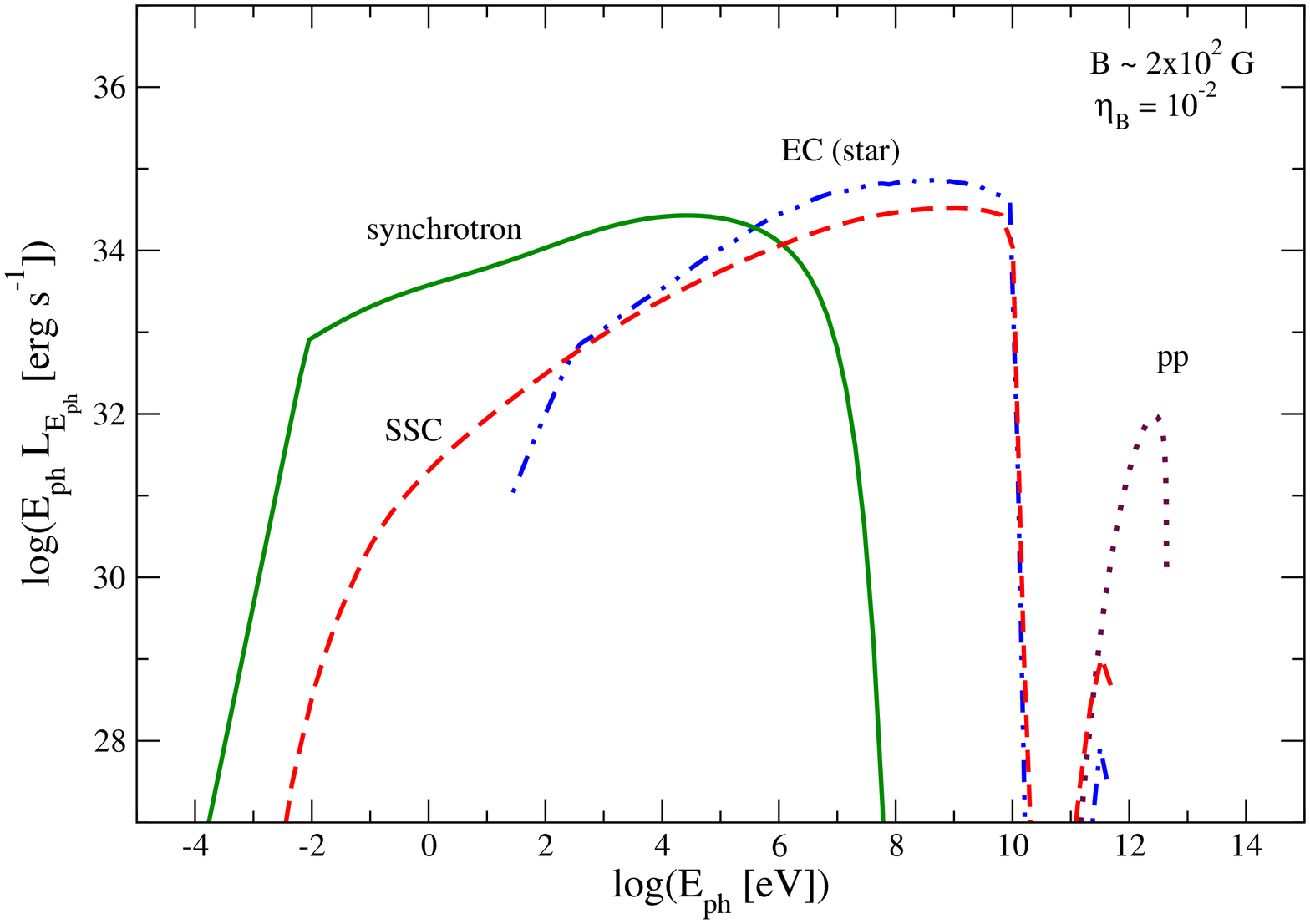}
\caption{Left: Timescales of the acceleration and lepton cooling losses 
(synchrotron, SSC, EC and relativistic Bremsstrahlung). Diffusion
and advection escape losses are also shown.    
Right: Spectral energy distribution produced by one clump of 
$R_{\rm c}^0 = 10^{10}$~cm interacting with the jet at $z_{\rm int} \sim a$. 
At energies $E_{\rm ph} \gtrsim 10$~GeV, the emission is strongly absorbed. 
Both (left and right) plots have been produced taking $B \sim 2\times10^2$~G.}
\label{losses_sed}
\end{figure}

The main radiative losses that affect the evolution of the non-thermal 
particles are synchrotron radiation, and synchrotron self-Compton (SSC) and
external Compton (EC) scattering. For EC cooling we have considered 
target photons provided by the WR star, that has a luminosity 
$L_{\star} \sim 10^{39}$~erg~s$^{-1}$ and a temperature $T_{\star} \sim 10^5$~K. 
The energy density of stellar photons at $z_{\rm int}$ is
$U_{\star} \sim 3\times10^{4}$~erg~cm$^{-3}$. At electron energies 
$E_e \gtrsim 25$~GeV, EC scattering with stellar photons of energy
$E_{\rm ph\star} \sim 3 K T_{\star}$ (where $K$ is the Boltzmann constant) 
occurs in the Klein-Nishina regime.
In addition to radiative losses, electrons are advected away from the 
emitter on a time $t_{\rm adv} \sim 4\,R_{\rm c}/v_{\rm j}$.
In Figure~\ref{losses_sed} (left), the electron cooling timescales are plotted 
together with the acceleration timescale. As seen in the figure, 
for $\eta_{\rm B} = 10^{-2}$ the maximum energy is 
determined by synchrotron losses, yielding $E_e^{\rm max} \sim 1$~TeV. 
To obtain the distribution $N_e$ of relativistic electrons 
we solve the kinetic equation~\cite{ginzburg} for a one-zone emitter,
taking into account the losses mentioned above, and also the  
synchrotron photons (as targets for SSC).  
The steady state is reached on a time $\ll t_{\rm sc}$.
$N_e(E_e)$ has a break at the energy $E_{\rm b} \sim 0.1$~GeV 
due to electron advection escape.

Protons can lose energy via $pp$ interactions 
in the bow-shock region, but diffusion losses dominate, 
constraining the maximum energy to 
$E_{p}^{\rm max} \sim 2.5\times10^2\sqrt{\eta_{\rm B}}$~TeV (for
$\eta_{\rm nt} = 0.1$).
The most energetic protons, $E_p \gtrsim 0.1\,E_{p}^{\rm max}$,
can diffuse up to the clump before escaping through advection, producing
gamma rays via $pp$ interactions.

\subsection{High-energy emission}

In the bow-shock region,  
using standard formulas~\cite{blumenthal} and $N_e$, we estimate the 
synchrotron, SSC and EC emission; relativistic Bremmstrahlung and $pp$
interactions are negligible there. 
In the clump, if energetic protons that arrive from the bow shock 
are not confined, they will  escape from the clump in a time 
$t_{\rm cl} \sim R_{\rm c}^0/c$ before radiating a significant part of 
their energy. Considering that the energy distribution of relativistic 
protons in the clump is 
$N_p \sim Q_p\,t_{\rm cl}$, we estimate the $pp$ emission following
the formulae given in~\cite{kelner}.  
In Figure~\ref{losses_sed} (right), we show the spectral energy distribution
(SED) composed by the most 
important radiative processes in the bow-shock region (synchrotron, SSC and
EC) and in the clump ($pp$). The synchrotron emission is self-absorbed at 
energies $E_{\rm ph} < 10^{-4}$~eV, and at photon energies 
$E_{\rm ph} \gtrsim 10$~GeV  absorption due to electron-positron pair creation
in the star photon field is relevant. 
The achieved luminosity at energies $\sim 0.1-10$~GeV is 
$L_{\rm EC} \sim 2\times10^{35}$~erg~s$^{-1}$.
Regarding to the $pp$ emission, the estimate obtained in this work
is a lower-limit. If the protons were confined in the clump, the radiated
emission would be larger.

The SED shown in Figure~\ref{losses_sed}~(right) has been 
calculated neglecting the expansion
of the clump inside the jet (see Figure~\ref{exp}, right). 
If we consider that after $t_{\rm sc}$ the cloud
expands up to a size much larger than $R_{\rm c}^0 = 10^{10}$~cm, 
the available non-thermal luminosity  will
be significantly larger and, as a consequence, the synchrotron and IC 
luminosity will be higher. On the other hand, the $pp$ emission 
produced in the expanded clump will be lower, because the clump density 
decreases faster than the increase of the clump surface: 
$n_{\rm c}/n_{\rm c}^0 \propto (R_{\rm c}^0/R_{\rm c})^3$. 

Finally, note that in Figure~\ref{losses_sed}~(right) we show the result 
of the interaction of only one clump with 
the jet, but many clumps could simultaneously interact with the jet.  
The non-thermal
radiation fluxes grow linearly with the number of clumps interacting with the 
jet, $N_{\rm cj}$, as long as too many clumps do not disrupt the jet. 
The corresponding SED will be also higher. 

\section{Discussion}

The total luminosity produced by jet-clump interactions depends on the 
number of clumps inside the jet, and on the size and density of each one.
The number of clumps in the jet can be estimated as 
$N_{\rm cj}\sim f_{\rm j}\,V_{\rm j}/V_{\rm c}$, where $f_{\rm j} \leq f$ is the 
filling factor
of clumps into the jet, and $V_{\rm j}$ and $V_{\rm c}$ are the jet and the 
clump volume, respectively. Considering $V_{\rm j}$ up to $z \sim 2\,a$, 
$V_{\rm c} \sim \pi {R_{\rm c}^0}^{3}$ (where $R_{\rm c}^0 = 10^{10}$~cm) 
and $f_{\rm j} = f = 0.01$, we obtain  
$N_{\rm cj} \sim 0.5$ in Cygnus~X-3. However, the real number of
clumps into the jet will be smaller than the previous estimate, 
because $f_{\rm j} < f$ as  a 
consequence of the clump destruction during the penetration process and
inside the jet due to RT and KH instabilities.
If $N_{\rm cj} < 1$, the emission
produced by the jet-clump interaction will be like a flare
with a duration determined by the lifetime of the clumps into the jet.
 
On the other hand, if we consider clumps with a radius $R_{\rm c}^0 = 10^9$~cm 
($\lesssim 0.01\,R_{\star}$), the number of clumps into the jet grows up to 
$\sim 10^4$.
Then, the simultaneous interaction of $\sim 10^4$ clumps with the jet will 
produce significantly more luminosity than the emission produced by the 
interaction of only one clump, that is $\sim 10^{34}$~erg~s$^{-1}$
 at $z_{\rm int} \sim a$. 
In that case, the final spectrum will be persistent, with a flickering due to
the continuous interactions of many small clumps. 

However, the jet can be disrupted by the too small many clumps, in particular
if the section of  clumps interacting within $\Delta z$ is comparable with the 
section of the jet: $\Sigma R_{\rm c}^2 \sim R_{\rm j}^2$.
This will be the case, neglecting clump expansion, for 
$R_{\rm c}^0 \lesssim 10^9$~cm. Note that periods of smaller clump size
may lead to jet disruption. Also, changes in the wind could lead to more
clumps inside the jet and more effective tapping of the jet luminosity.

\end{document}